\documentclass[pre,preprint,showpacs,floatfix]{revtex4-1}
\usepackage{graphicx}
\usepackage{amsmath}
\usepackage{amsfonts}
\usepackage{amssymb}
\usepackage{hyperref}
\usepackage{bm}
\DeclareGraphicsExtensions{.jpg,.pdf}
\begin{document}
\title{Improving the Frequency Precision of Oscillators by Synchronization}
\author{M. C. Cross}
\affiliation{Department of Physics, California Institute of
Technology, Pasadena CA 91125}

\date{\today}

\begin{abstract}
Improving the frequency precision by synchronizing a lattice of oscillators is studied in the phase reduction limit. For the most commonly studied case of purely dissipative phase coupling (the Kuramoto model) I confirm that the frequency precision of $N$ entrained oscillators perturbed by independent noise sources is improved by a factor $N$ as expected from simple averaging arguments. In the presence of reactive coupling, such as will typically be the case for non-dissipatively coupled oscillators based on high-Q resonators, the synchronized state consists of target like waves radiating from a local source which is a region of higher frequency oscillators. In this state all the oscillators evolve with the same frequency, however I show that the improvement of the frequency precision is independent of $N$ for large $N$, but instead depends on the disorder and reflects the dependence of the frequency of the synchronized state on just those oscillators in the source region of the waves.	

\end{abstract}

\maketitle

\section{Introduction}

Oscillators -- devices producing a periodic signal at a frequency determined by the characteristics of the device and not by an external clock -- play a crucial role in much of modern technology, for example in timekeeping (quartz crystal watches), communication (frequency references for mixing down radio frequency signals), and sensors. A key characteristic is the intrinsic frequency precision of the device, which can be quantified in terms of the line width of the oscillating signal, or in more detail by the spectral density of the signal in the frequency domain, or by the Allan deviation in the time domain. This is the fundamental issue, broadly common to all oscillators, considered in the present paper. There are other important practical characteristics, including the robustness of the frequency to environmental perturbations such as vibrations and temperature fluctuations, that are more dependent on the details of the device implementation; these are not considered here.

Unlike a resonator driven by an external oscillating signal, where the line width of the spectral response is determined by the dissipation in the resonator, the line width of an oscillator is only nonzero in the presence of noise. An oscillator is mathematically described by a limit cycle in the phase space of dynamical variables, and the line width of the signal corresponding to a limit cycle is zero. Dissipation serves to relax the system to the limit cycle, which is itself determined by a balance of energy injection and dissipation, and does not broaden the spectral line. The spectral line is broadened only if there is some stochastic influence that causes the phase space trajectory to fluctuate away from the limit cycle. Thus the frequency imprecision (line width) is due to noise\cite{Lax67}.

One way to improve the frequency precision of oscillators that has been suggested in various scientific disciplines\cite{ChangCao97a, NeedlemanTiesinga07}, is to sum the signal from a number $N$ similar oscillators: in this case, if the noise sources are uncorrelated over the individual oscillators, simple averaging suggests that the effective noise intensity will be reduced by a factor $1/N$. Of course, due to fabrication imperfection the isolated oscillators will have slightly different frequencies, and so simply averaging the summed signal from the individual oscillators will also tend to average the signal to zero. In addition, the line width of the reduced intensity signal will reflect the frequency dispersion of the devices. If however some coupling is introduced between the oscillators, they may become \emph{synchronized} to a state in which all the oscillators are entrained to run at a single common frequency\cite{PikovskyRosenblaum01}. In this case, the averaging argument would suggest a $1/N$ reduction in the effective noise, and so a factor of $N$ enhancement in the frequency precision.

In this paper I investigate the improvement of frequency precision due to synchronization in canonical models of the phenomenon. I focus on lattices of oscillators with nearest neighbor coupling. I confirm that the factor  $N$ improvement applies exactly in the case of purely dissipative coupling between the phases of the oscillators. However, this scaling breaks down if there is in addition a reactive (non-dissipative) component of the coupling. Examples of reactive coupling are a displacement, rather than velocity, coupling for arrays of mechanical oscillators\cite{,CrossZumdieck04, CrossRogers06} or trapped ions\cite{LeeCross11}. In this case the improvement factor becomes independent of $N$ for large $N$, but depends on the amount of disorder. This may lead to much poorer frequency precision than anticipated from the na\"ive averaging argument.

The main focus of the paper is on d-dimensional lattices of oscillators with nearest neighbor coupling. I briefly discuss the extension to longer range coupling and to complex networks. These results should be relevant to questions of the precision of synchronized oscillations in biological contexts.

\section{Model}
The model I consider is $N$ nearest neighbor coupled phase oscillators\cite{Winfree67,Kuramoto75,KuramotoBook, PikovskyRosenblaum01, AcebronBonilla05}
\begin{equation}
\dot\theta_{i}=\omega_{i}+\sum_{\text{nn }j}\Gamma(\theta_{j}-\theta_{i})+\xi_{i}(t),\qquad i=1,\ldots,N.\label{Phase}
\end{equation}
Here $\omega_{i}$ are the frequencies of the individual oscillators, which are assumed to be independent random variables taken from a distribution $g(\omega)$ with width $\sigma$\footnote{For distributions with finite variance, $\sigma$ can be defined as the standard deviation. For distributions without a variance, such as a Lorentzian, some other appropriate characterization of the width may be chosen.}. The term $\xi_{i}(t)$ represents the noise acting on the $i$th oscillator. I will assume white noise, but the results are easily generalized to colored noise, and also to noise that depends on the phase $\theta_{i}$. An important assumption is that the noise is \emph{uncorrelated} between different oscillators. Thus
\begin{equation}
\langle\xi_{i}(t)\xi_{j}(t')\rangle=c(t-t')\delta_{ij},
\end{equation}
where for white noise
\begin{equation}
c(t-t')=f\delta(t-t'),
\end{equation}
with $f$ the individual oscillator noise strength, taken to be the same for all oscillators, since I am imagining a system where the oscillators are designed to be as similar as possible.

The coupling between the oscillators is given by the function $\Gamma$, a $2\pi$ periodic function of the phase differences of nearest neighbor oscillators. A commonly used model\cite{Winfree67, Kuramoto75} is given by the coupling function
\begin{equation}
\Gamma(\phi)=\sin\phi.\label{Dissipative coupling}
\end{equation}
Any parameter $K$ multiplying $\sin\phi$ and giving the strength of the coupling may be scaled to unity by rescaling time and frequencies.
Thus the only parameters defining the behavior of the system are the distribution of the frequencies $\omega_{i}$, and in particular the width $\sigma$ of the distribution after this rescaling (i.e., the width of the frequency distribution relative to the coupling strength).

The coupling function Eq.~(\ref{Dissipative coupling}) is antisymmetric $\Gamma(-\phi)=-\Gamma(\phi)$, and equation (\ref{Phase}) is purely dissipative\cite{TopajPikovsky02}. A more general coupling function
\begin{equation}
\Gamma(\phi)=\sin\phi+\gamma(1-\cos\phi)\label{General coupling}
\end{equation}
breaks this symmetry for nonzero $\gamma$, and includes non-dissipative, propagating effects\cite{KuramotoBook,SakaguchiShinomoto88}. I will use this model to study the effect of reactive coupling. Without loss of generality, I take $\gamma>0$.

For small disorder, the phase difference between nearest neighbor oscillators will be small. A convenient approximation for $\Gamma$ good for small phase differences is\cite{BlasiusTonjes05}
\begin{equation}
\Gamma(\phi)\simeq\gamma^{-1}(e^{\gamma\phi}-1),\quad|\phi|,|\gamma\phi|\ll 1.\label{Small phase difference coupling}
\end{equation}

\section{Synchronization}
I first describe the behavior predicted by Eqs.~(\ref{Phase}-\ref{General coupling}) in the absence of noise.
For sufficiently weak disorder (small $\sigma$) and for a finite number of oscillators, the oscillations described by Eqs.~(1-3) with$f=0$ become entrained in the sense that all the phases advance at the same constant rate
\begin{equation}
\dot\theta_{i}=\Omega.
\end{equation}
The solution to these equations can be written
\begin{equation}
\theta_{i}(t)=\theta_{i}^{(s)}+\Theta(t),\label{Limit cycle}
\end{equation}
with $\theta_{i}^{(s)}$ a fixed point solution in the rotating frame, and $\Theta$ the phase of the collective limit cycle given by
\begin{equation}
\Theta(t)=\Omega t+\Theta_{0},\label{theta(t)}
\end{equation}
with $\Theta_{0}$ an arbitrary constant. $\theta_{i}^{(s)}$ and $\Omega$ are given by solving
\begin{equation}
\Omega=\omega_{i}+\sum_{\text{nn }j}\Gamma(\theta_{j}^{(s)}-\theta_{i}^{(s)}),\label{Fixed point}
\end{equation}
when solutions exist.
The behavior in the thermodynamic limit $N\to\infty$ for different lattice dimensions $d$, and the critical value of $\sigma$ for the onset of this entrained state and its dependence on system size, lattice dimension, frequency distribution etc., are subtle questions that have not been fully answered. But for the practical case of a finite number of oscillators, we expect that an entrained (fully frequency locked) state will exist for sufficiently small $\sigma$\footnote{There may be multistability, so that other non-entrained states may exist for the same parameter values: I will only consider the entrained state.}. Such a state is a limit cycle of the system of oscillators with frequency $\Omega$. By moving to a rotating frame, $\theta_{i}(t)\to\theta_{i}(t)-\Omega t$, the entrained state becomes a fixed point, simplifying the subsequent analysis.

The nature of the synchronized state depends sensitively on whether the coupling is purely dissipative ($\gamma=0$) or also contains a reactive component ($\gamma \ne 0$). For purely dissipative coupling the interactions cancel when summed over a block of oscillators. This means that the frequency of the entrained state is the mean $\bar\omega$ of the oscillator frequencies. For a one-dimensional lattice, the individual phases $\theta_{i}^{(s)}$ are then given by $\theta_{i+1}^{(s)}-\theta_{i}^{(s)}=-\sin^{-1}X_{i}$ with $X_{i}=\sum_{j=1}^{i}(\omega_{j}-\bar\omega)$\cite{StrogatzMirollo88b}. The accumulated randomness $X_{i}$ performs a random walk as a function of the lattice index $i$, and so the strain $\theta_{i+1}^{(s)}-\theta_{i}^{(s)}$ also varies with $i$ roughly as a random walk. The break down of the synchronized state as the disorder or system size increases occurs when the excursion of $X_{i}$ exceeds unity (remember the coupling strength is scaled to one). This occurs for $\sigma\sim N^{-1/2}$. On the other hand, for $\gamma \ne 0$, the interaction terms summed over a block of oscillators do not cancel, and the same arguments cannot be made. For $\gamma > 0$ it is found that the entrained state takes the from of quasi-regular waves of some average wave length $\lambda$ propagating away from a unique source in the system, located at a cluster of higher frequency oscillators\cite{SakaguchiShinomoto88,BlasiusTonjes05}. The derivation of this result is described in more detail below.
In this wave state the frequencies of all the oscillators remain entrained, even though the phases vary by more than $2\pi$ over the system for $\lambda < N$, and by many factors of $2\pi$ for $\lambda\ll N$. In two dimensional lattices, roughly circular ``target'' waves are found for $\gamma \ne 0$, with the waves propagating away from a source with location again given by a core region of higher frequency oscillators\cite{SakaguchiShinomoto88,BlasiusTonjes05}.

\section{Frequency precision and noise}

The noise terms in Eqs.~(\ref{Phase}) will lead to deviations of the solution from the limit cycle Eqs.~(\ref{Limit cycle},\ref{theta(t)}) and so to a broadening of the spectral lines of the output signal from the oscillator. The full noise spectrum depends on a complete solution of Eqs.~(\ref{Phase}). However, for frequency offsets from the no-noise peaks in the power spectrum that are small compared with the relaxation rates onto the limit cycle, the effects of the noise can be reduced to a single stochastic equation for the limit cycle phase $\Theta$ that gives the collective behavior of the entrained oscillators. This result has been derived for a general limit cycle by a number of authors using a variety of formalisms\cite{Lax67,Kaertner90, DemirMehrotra00,GoldobinTeramae10}. The key idea is that a change in $\Theta$ corresponds to a time translation, and so gives an equally good limit cycle solution: thus a perturbation to $\Theta$  does not decay, and this represents a zero-eigenvalue mode of the linear stability analysis\footnote{In the general case the stability analysis would be a Floquet analysis of a periodic state;  in the present case this can be reduced to a stability analysis of the fixed point in the rotating frame.}.  The remaining eigenvalues of the stability analysis will be negative, corresponding to exponential decay onto the limit cycle. For time scales longer than these relaxation times, it is only the projection of the noise along the zero eigenvalue eigenvector, that is important: the other fluctuation components will have decayed away.

For the white noise sources 
considered here, this stochastic equation for the phase is simply\cite{GoldobinTeramae10}
\begin{equation}
\dot\Theta(t)=\bar\Omega+\Xi(t),\label{Stochastic phase}
\end{equation}
with
\begin{equation}
\langle\Xi(t)\Xi(t')\rangle=F\delta(t-t'),\label{Phase noise}
\end{equation}
with $F$ the noise strength resulting from the projection and $\bar\Omega$ the limit cycle frequency which is $\Omega$ with an $O(F^{2})$ correction. The solution to Eq.~(\ref{Stochastic phase}) is a drift of the mean phase at the rate $\bar\Omega$
\begin{equation}
\langle\Theta(t)\rangle=\bar\Omega t,
\end{equation}
together with phase diffusion
\begin{equation}
\langle(\Theta(t)-\bar\Omega t)^{2}\rangle=Ft.
\end{equation}
An output signal from the oscillator such as $X=\cos\Theta(t)$ will have a power spectrum consisting of a Lorentzian peak centered at
$\bar\Omega$ (and, for more general signals, the harmonics)
\begin{equation}
S_{XX}(\omega)=\frac{S_{0}}{2\pi}\frac{F}{(\omega-\bar\Omega)^{2}+\tfrac{1}{4}F^{2}},
\end{equation}
with $S_{0}$ the spectral weight of the delta-function peak in the spectrum of the no-noise oscillator\cite{DemirMehrotra00}. The width of the spectral peak is therefore equal to the phase noise strength $F$. Thus the tails of the spectrum away from the peaks decay as $\omega^{-2}$; this is the white-noise component of the Leeson noise spectrum for oscillators\cite{Leeson66}. Other noise spectra will lead to different power law tails.

The relationship of the effective noise strength $F$ acting on the collective phase $\Theta$ to the the strength of the noise $f$ acting on each individual oscillator is given by projecting the individual noise components $\xi_{i}(t)$ along the phase variable $\Theta$. Denoting the phases $\theta_{i}$ by the vector $\bm\theta$, the tangent vector to the limit cycle (the zero-eigenvalue eigenvector) is given by $\mathbf e_{0}=(1,1,1\ldots,1)$, choosing a convenient normalization so that the phase shifts corresponding to a time translation $\Delta t$ are $\delta\theta_{i}= e_{0,i}\Delta t$. Using the general results of refs.~\onlinecite {Kaertner90, DemirMehrotra00,GoldobinTeramae10}, or the simpler analysis for the present case sketched in Appendix \ref{Appendix}, the relationship is
\begin{equation}
F=\frac{\mathbf e_{0}^{\ \dagger}\cdot \mathbf e_{0}^{\ \dagger}}{(\mathbf e_{0}^{\ \dagger}\cdot\mathbf e_{0})^{2}}\ f,\label{Noise reduction}
\end{equation}
with $\mathbf e_{0}^{\ \dagger}$ the zero-eigenvalue adjoint eigenvector. Thus finding the broadening of the line due to the noise is reduced to calculating the adjoint eigenvector $\mathbf e_{0}^{\ \dagger}$.

The Jacobean matrix $\mathbf{J}$ yielding the linear stability analysis of the phase dynamics about the fixed point phases $\theta_{i}^{(s)}$  defining the limit cycle is
\begin{subequations}
\label{Jacobean}
\begin{align}
J_{ij}&=\Gamma'(\theta_{j}^{(s)}-\theta_{i}^{(s)})\text{ for i,j nearest neighbors},\\
J_{ii}&=-\sum_{\text{nn }j}\Gamma'(\theta_{j}^{(s)}-\theta_{i}^{(s)}),
\end{align}
with other elements zero. The vectors $\mathbf e_{0},\mathbf e_{0}^{\ \dagger}$ are defined by
\begin{align}
&\mathbf{J}\cdot\mathbf{e}_{0}=0,\\
&\mathbf{J}^{\dagger}\cdot\mathbf{e}_{0}^{\ \dagger}=0,
\end{align}
\end{subequations}
with $J^{\dagger}_{\ ij}=J_{ji}$.

Note that I am treating noise perturbatively in the small noise limit and for a finite system: in this case the result is given by just the effect on the overall phase of the synchronized state, which is the zero-mode of the system. I am not considering modifications to the synchronized state due to the noise, such as changes in values of the critical disorder for synchronization, or changes in the nature of the synchronized state. In a finite system there will be barriers to such fluctuations, and their rates will vary with the noise strength $f$ as $e^{-\Delta/f}$ with $\Delta$ some number depending on the states considered. These fluctuations can therefore be ignored for small enough $f$. As the number of oscillators tends to infinity, some barriers will become very small, and the synchronized state may be significantly changed or even eliminated by the addition of noise\cite{AcebronBonilla05}, as for phase transitions in equilibrium systems at finite temperature.

\section{Dissipative coupling}
For purely dissipative coupling $\gamma=0$, the Jacobean $\mathbf J$ is symmetric and so the adjoint eigenvector is equal to the forward eigenvector which is the tangent vector defined by an infinitesimal time translation
\begin{equation}
\mathbf e_{0}^{\ \dagger}=\mathbf e_{0}=(1,1,1\ldots,1).
\end{equation}
This result is true for any antisymmetric coupling, and is not restricted to the nearest neighbor model. This immediately gives the result for the effective noise strength
\begin{equation}
F=N^{-1}f,\label{Noise: dissipative coupling}
\end{equation}
so that the frequency precision of the entrained state is enhanced by the factor $N$. Note that the enhancement does \emph{not} depend on the degree of phase alignment quantified by the magnitude of the order parameter $\Psi=N^{-1}\sum_{j}e^{i\theta_{j}^{(s)}}$, which may be less than unity (i.e., phases not fully aligned) even in the entrained state. The result Eq.~(\ref{Noise: dissipative coupling}) has been obtained previously\cite{ChangCao97a,NeedlemanTiesinga07}, although I believe the present derivation is more systematic, since it does not assume that the effect of the noise on the collective phase remains small at long times.

\section{Dissipative plus reactive coupling}

For general coupling the Jacobean is not symmetric, and there is no obvious relationship between $\mathbf e_{0}^{\ \dagger}$ and $\mathbf e_{0}$ in general. Physically, in situations where the entrained state consists of waves emanating from a source region of higher frequency oscillators, we might expect the frequency precision to be determined by fluctuations of only those oscillators in the core region that fix the frequency of the waves. This means that the reduction of the effective noise by averaging is only over this core region of oscillators, giving a poorer improvement of the frequency precision. I first demonstrate this result for a simpler ``one-way'' coupling function introduced by Blasius and Tonjes\cite{BlasiusTonjes05} for which analytic solution is possible. I then derive the result for the general coupling function Eq.~(\ref{General coupling}) assuming the disorder is small enough so that the phase difference between all nearest neighbor oscillators is small, in which case the approximation Eq.~(\ref{Small phase difference coupling}) may be used. The result depends on the mapping\cite{SakaguchiShinomoto88,BlasiusTonjes05} of the solution for the entrained state onto the Anderson localization problem\cite{Anderson58}, and the known properties of the localized states in this problem\cite{LeeRamakrishnan85,Jensen98}, together with a relationship between $\mathbf e_{0}^{\ \dagger}$ and the localized states that I demonstrate. I also investigate one and two dimensional lattices numerically.

\subsection{One-way coupling}
\label{Sec: One-way}

\begin{figure}[tbh]
\begin{center}
\includegraphics[width=0.46\columnwidth]{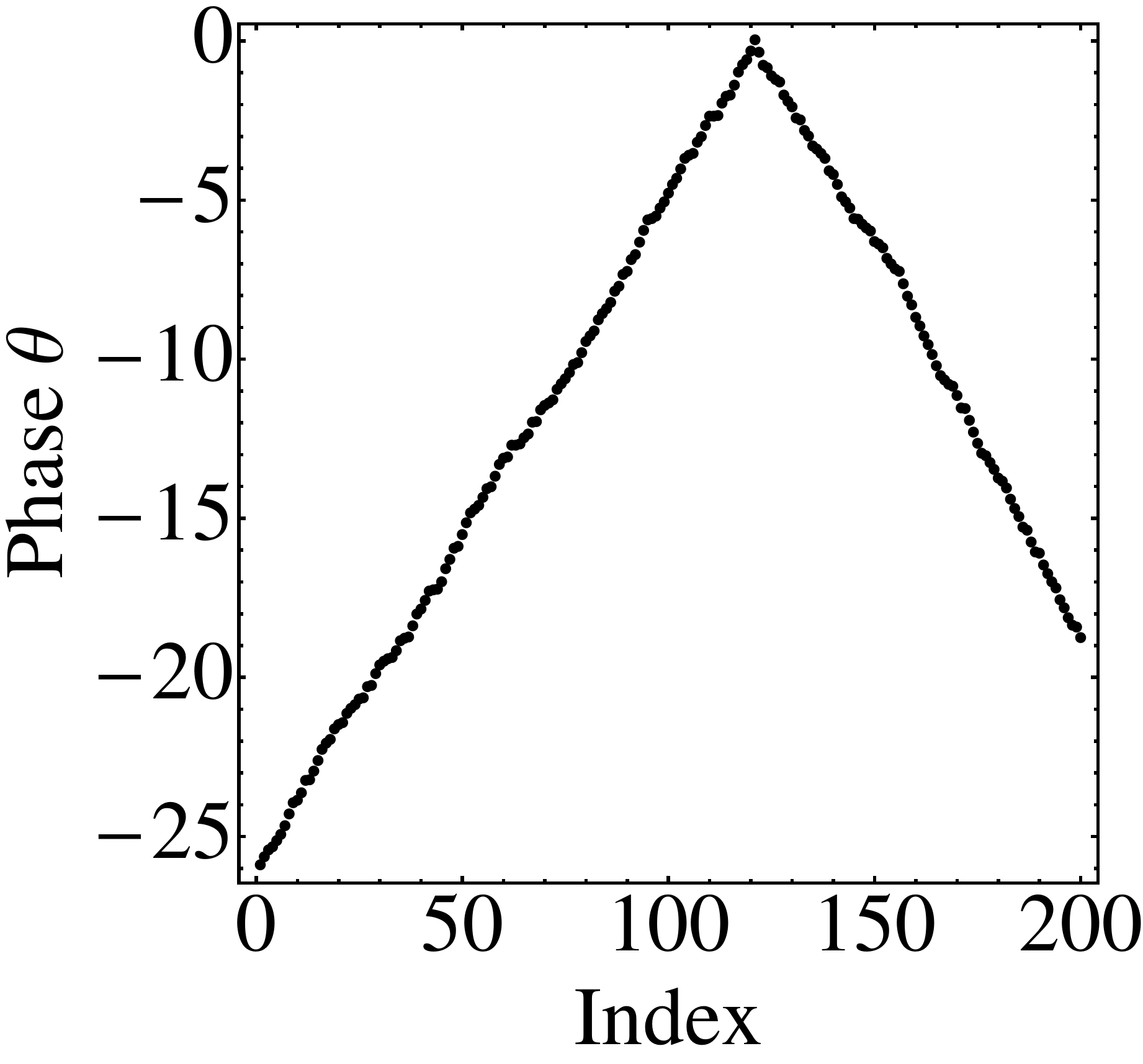}
\hspace{0.05\columnwidth}
\includegraphics[width=0.46\columnwidth]{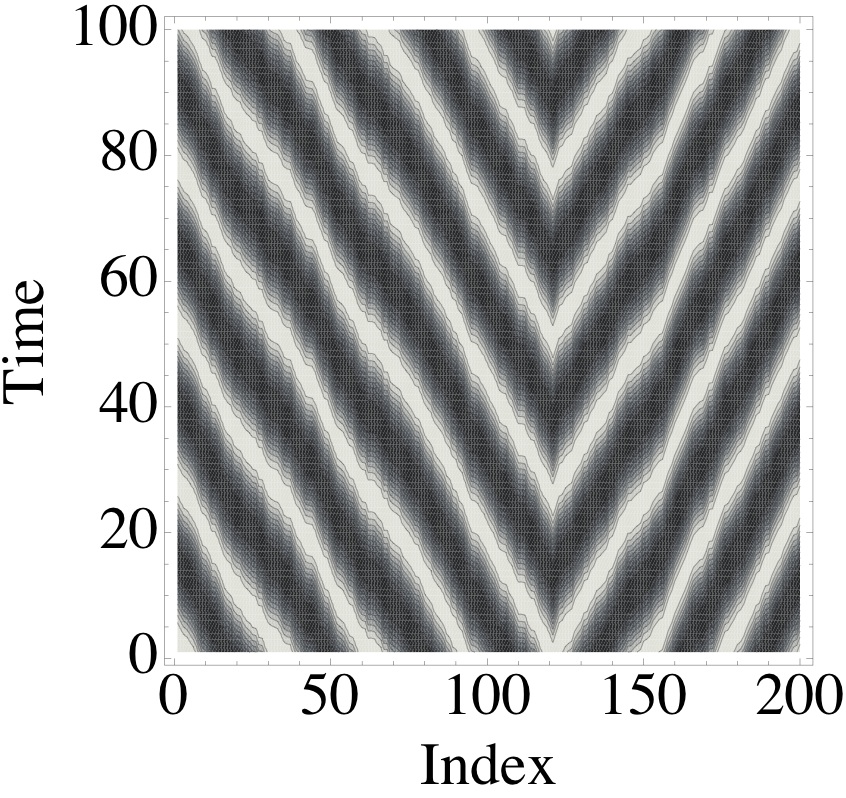}
\caption{\label{Fig: one sided}Entrained state for the one sided model Eq.~(\ref{One sided model}) for 200 oscillators. Left panel: oscillator phase as a function of lattice site with $\theta_{m}$ set to 0; right panel: gray scale plot of $\cos\theta_{i}$ as a function of time. The oscillator frequencies were taken from a uniform frequency distribution with width $\sigma=0.5$ and mean zero, and $\gamma=1$.}
\end{center}
\end{figure}

Blasius and Tonjes\cite{BlasiusTonjes05} proposed a simple, exactly soluble model of a one dimensional lattice, with a nearest neighbor coupling function such that the phase of oscillator $i$ is only influenced by its neighbors if their phases are ahead (all phase differences are assumed to be small so that this notion makes sense). I use the example
\begin{equation}
\Gamma(\phi)=\left\{\begin{array}{cc}\gamma(e^{\gamma\phi}-1) & \text{ for }\phi>0, \\0 & \text{ for }\phi <0.\end{array}\right.\label{One sided model}
\end{equation}
The entrained solution is given by $\Omega=\omega_{m}$ with $m$ the index of the largest frequency in the lattice, and then the fixed point solution $\bm\theta^{(s)}$ is constructed iteratively from $\theta^{(s)}_{m}$ using
\begin{equation}
\theta^{(s)}_{i}=\left\{\begin{array}{cc}\theta^{(s)}_{i-1}-\gamma^{-1}\ln[1+\gamma(\omega_{m}-\omega_{i})] & \text{ for }i>m, \\
\theta^{(s)}_{i+1}-\gamma^{-1}\ln[1+\gamma(\omega_{m}-\omega_{i})]  & \text{ for }i<m.\end{array}\right.
\end{equation}
The value chosen for $\theta^{(s)}_{m}$ sets the overall phase $\Theta$.
An example of the entrained state for 200 oscillators in a 1d lattice is given in Fig.~\ref{Fig: one sided}, showing waves emanating from the oscillator with maximum frequency at $m=121$.

It is easy to see for this coupling function that the zero-eigenvalue adjoint eigenvector is
\begin{equation}
e_{0,i}^{\ \dagger}=\delta_{im},
\end{equation}
corresponding to the fact the phase $\theta_{m}$ is not coupled to either neighbor, since $\theta_{m}>\theta_{m\pm 1}$. Thus the effective noise is given by $F=f$, and there is \emph{no} improvement of the frequency precision, even though all the oscillators are entrained. This is because the single oscillator with maximum frequency determines the entrained frequency, and therefore the entrained frequency is as sensitive to noise as this single oscillator.

\subsection{General coupling}
\label{Sec: General}

\begin{figure}[tbh]
\begin{center}
\includegraphics[width=0.46\columnwidth]{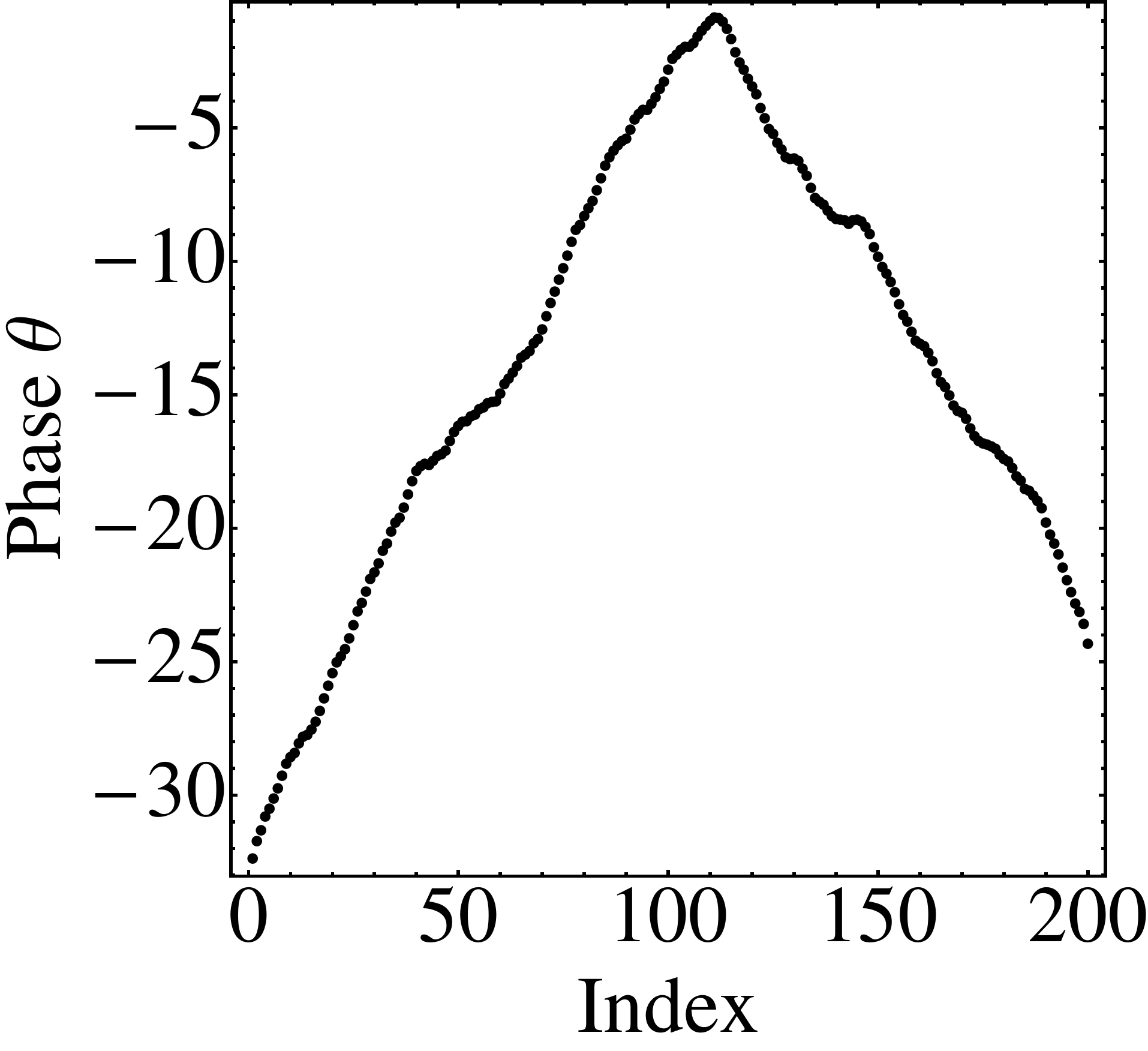}
\hspace{0.05\columnwidth}
\includegraphics[width=0.46\columnwidth]{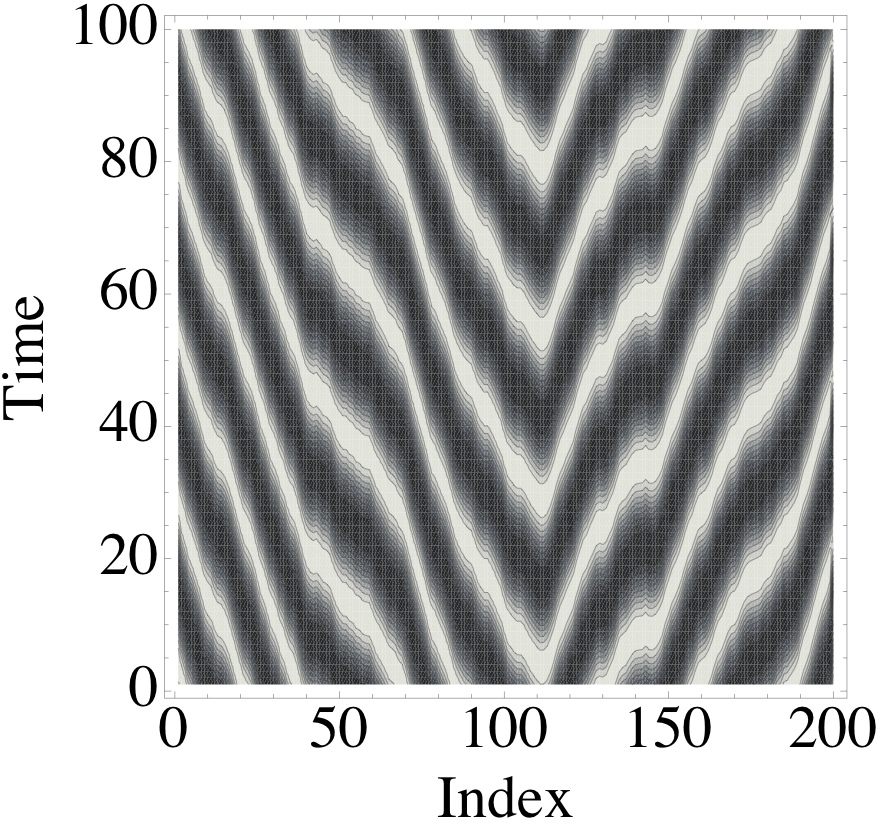}
\caption{\label{Fig: general}Entrained state for the general model Eq.~(\ref{General coupling}) for a chain of 200 oscillators with nearest neighbor coupling, using the small phase difference approximation Eq.~(\ref{Small phase difference coupling}). Left panel: oscillator phase as a function of lattice site; right panel: gray scale plot of $\cos\theta_{i}$ as a function of time. The oscillator frequencies were taken from a uniform frequency distribution with width $\sigma=0.5$ and mean zero, and $\gamma=1$.}
\end{center}
\end{figure}

I now consider the case of general coupling Eq.~(\ref{General coupling}) in the limit of small enough disorder so that the small phase difference approximation Eq.~(\ref{Small phase difference coupling}) approximation may be used. In this case, Blasius and Tonjes\cite{BlasiusTonjes05} showed that the Cole-Hopf transformation $\theta_{i}=\gamma^{-1}\ln q_{i}$ maps the problem for the entrained state onto the linear problem
\begin{equation}
\dot q_{i}=Eq_{i}=\gamma\omega_{i}q_{i}+\sum_{\text{nn }j}(q_{j}-q_{i}),\label{Linear}
\end{equation}
with the eigenvalue $E=\gamma\Omega$.
This is equivalent to the tight binding model for a quantum particle on a random lattice, and the properties of the solution can be extracted in analogy with Anderson localization\cite{Anderson58}. At long times the solution $\mathbf q(t)=\mathbf q^{\text{max}}e^{E_{\text{max}}t}$ corresponding to the largest eigenvalue $E_{\text{max}}$ will dominate. This gives the entrained state
\begin{equation}
\theta_{i}^{(s)}=\gamma^{-1}\ln q^{\text{max}}_{i},\label{theta q}
\end{equation}
with frequency $\Omega=E_{\text{max}}/\gamma$.
Anderson localization theory shows that $\mathbf q^{\text{max}}$ may be chosen positive, and it has the form of an exponentially  localized state centered on a region of the lattice with a concentration of larger frequency oscillators. The exponential localization of $\mathbf q^{\text{max}}$ corresponds to a roughy linear phase profile, again leading to waves propagating from a source, as shown in Fig.~\ref{Fig: general}.

I now analyze the frequency precision based on the properties of the solution $\mathbf q^{\text{max}}$ known from studies of the Anderson problem.

\begin{figure}[tbh]
\begin{center}
\includegraphics[width=0.46\columnwidth]{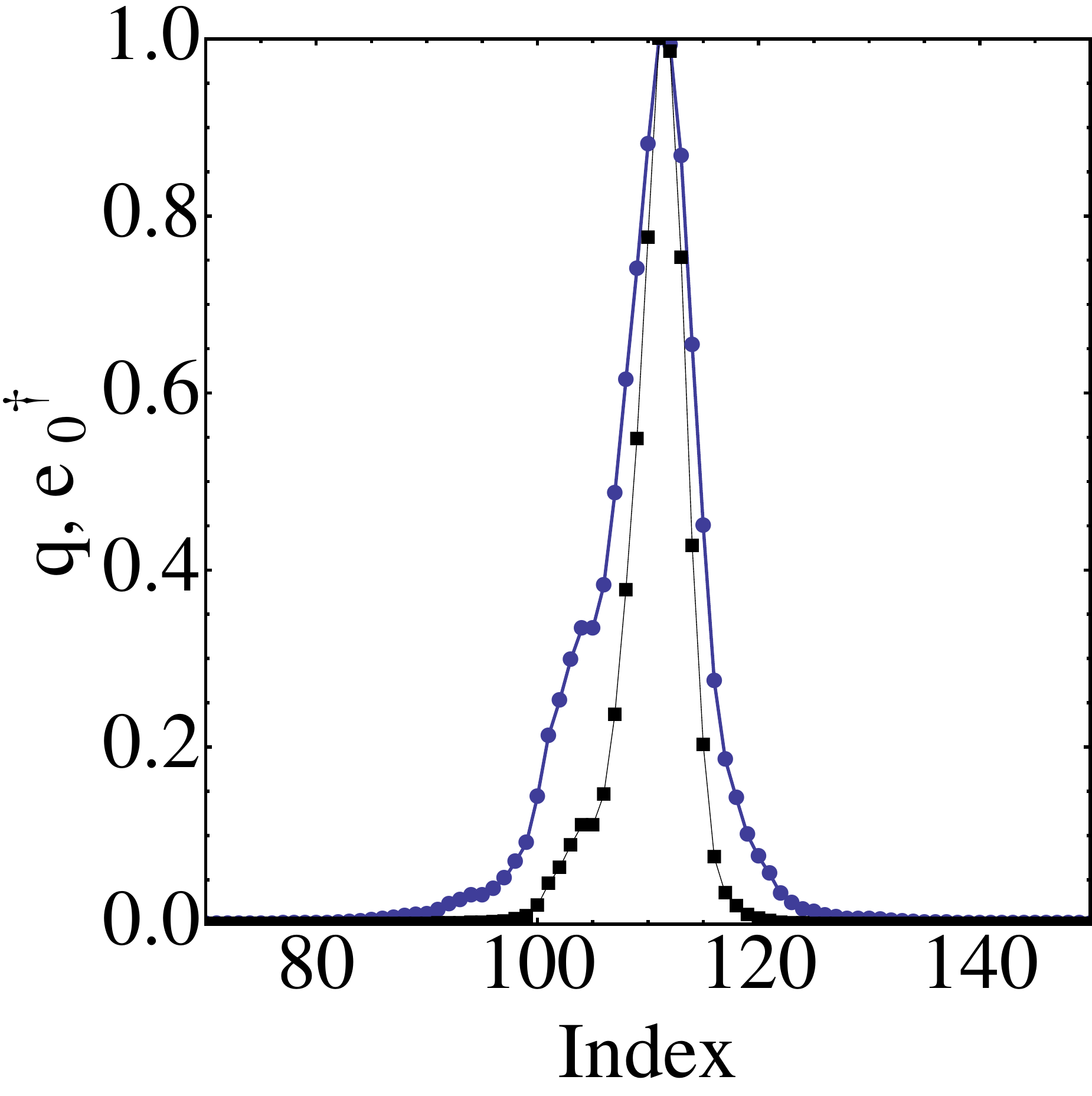}
\hspace{0.05\columnwidth}
\includegraphics[width=0.46\columnwidth]{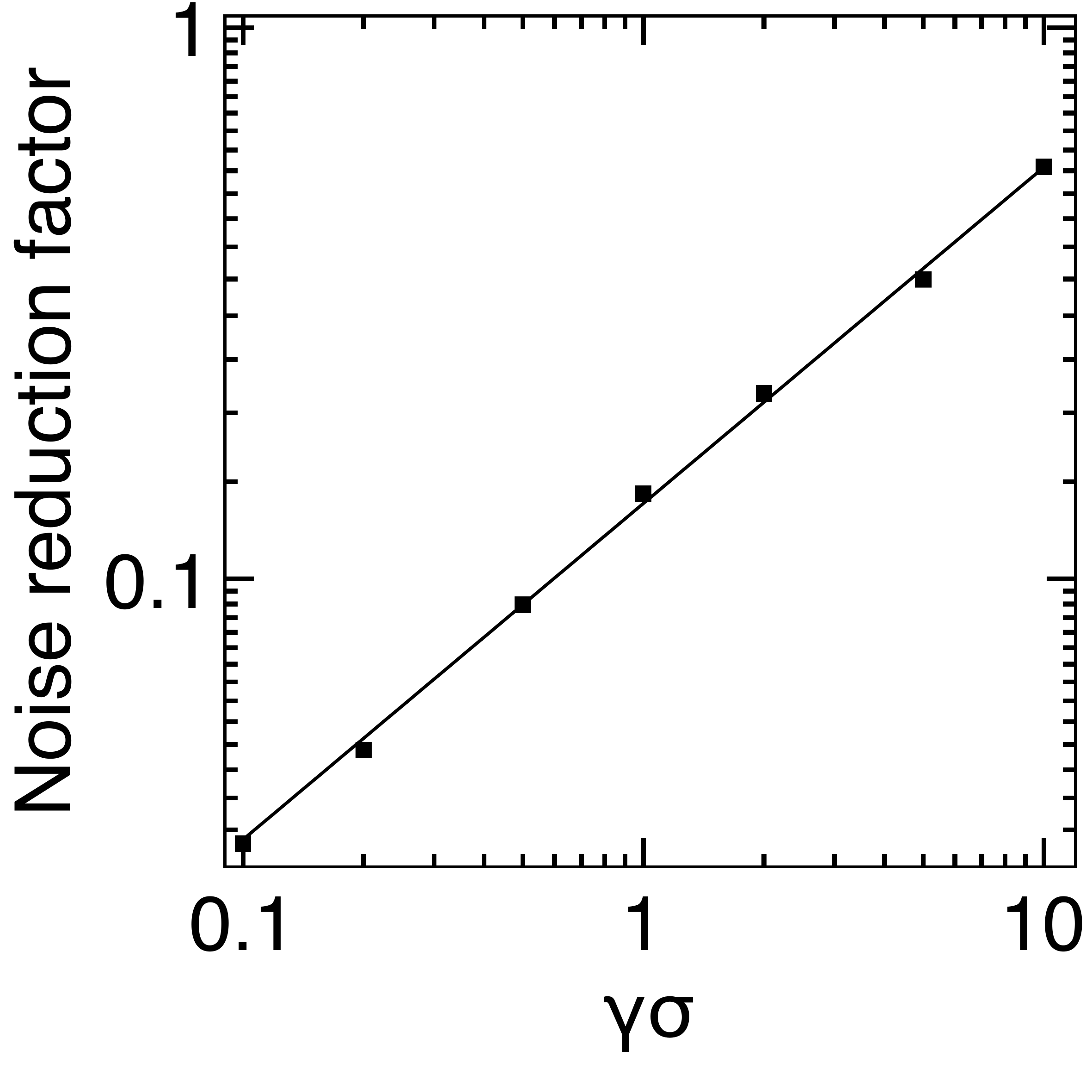}
\caption{\label{Fig: general qe0}Left panel: Localized solution $\mathbf q^{\text{max}}$ (blue circles) and zero-eigenvalue adjoint eigenvector $\mathbf e^{\ \dagger}_{0}$ (black squares) for the system of Fig.~\ref{Fig: general}. (The normalizations are chosen for the plot so that the largest element of each vector is 1.) Only the portion of the $L=200$ lattice where the elements have appreciable size is shown. Right panel: scaling of the noise reduction factor $F/f$ deduced from Eq.~(\ref{F/f}) with $\gamma\sigma$. Each point is the average of 1000 realizations of the random lattice of frequencies.}
\end{center}
\end{figure}

\subsubsection{One-dimensional lattice}
For a one-dimensional lattice with nearest neighbor coupling Eq.~(\ref{Small phase difference coupling}), the Jacobean matrix Eq.\~(\ref{Jacobean}) for the stability analysis of the fixed point solution $\theta^{(s)}_{i}$ is the tridiagonal matrix with elements
\begin{align}
J_{ii\pm 1}&=e^{\gamma(\theta^{(s)}_{i\pm 1}-\theta^{(s)}_{i})}=q^{\text{max}}_{i\pm 1}/q^{\text{max}}_{i},\\
J_{ii}&=-J_{i,i+1}-J_{i,i-1},
\end{align}
except for the first and last rows corresponding to the end oscillators which only have one neighbor
\begin{equation}
J_{12}=-J_{11}=q_{2}^{\text{max}}/q_{1}^{\text{max}}, J_{NN-1}=-J_{NN}=q_{N-1}^{\text{max}}/q_{N}^{\text{max}},
\end{equation}
and all other elements zero. It is easily checked that $(1,1,1\ldots,1)$ is indeed the zero-eigenvalue eigenvector.
The adjoint matrix has off diagonal elements
\begin{equation}
J^{\dagger}_{\ ii\pm 1}=q^{\text{max}}_{i}/q^{\text{max}}_{i\pm 1}
\end{equation}
except for the first and last rows for which
\begin{equation}
J^{\ \dagger}_{12}=q_{1}^{\text{max}}/q_{2}^{\text{max}},\quad J^{\ \dagger}_{NN-1}=q_{N}^{\text{max}}/q_{N-1}^{\text{max}},
\end{equation}
and diagonal elements
\begin{equation}
J^{\dagger}_{\ ii}=J_{ii},
\end{equation}
with all other elements zero.
The key result is that the (unnormalized) adjoint eigenvector can be found explicitly
\begin{equation}
e_{0,i}^{\ \dagger}=(q^{\text{max}}_{i})^{2},\label{adjoint}
\end{equation}
as can be confirmed by direct substitution. This simple result follows from the quotient form of the Jacobean matrix elements for the special form of the interaction Eq.~(\ref{Small phase difference coupling}).
An example of the vector $\mathbf q^{\text{max}}$ and the adjoint eigenvector $\mathbf e^{\ \dagger}_{0}$ for the system of Fig.~\ref{Fig: general} is shown in Fig.~\ref{Fig: general qe0}.

The noise reduction factor $F/f$, Eq.~(\ref{Noise reduction}), is given by
\begin{equation}
\frac{F}{f}=\frac{\sum_{i}(q_{i}^{\text{max}})^{4}}{\left[\sum_{i}\left(q_{i}^{\text{max}}\right)^{2}\right]^{2}}.\label{F/f}
\end{equation}
This equation directly relates the improvement in frequency precision to the solution of the linear Anderson problem Eq.~(\ref{Linear}).
The expression Eq.~(\ref{F/f}) for the noise reduction is the \emph{inverse participation ratio} $p^{-1}$ of the vector $\mathbf q^{\text{max}}$ of the linear localization problem. This can be used to define the radius of the localized state $r\equiv p/2$. Thus I find that the noise reduction factor is given by the size of the source of the waves, rather than by the total number of oscillators, giving an improvement in frequency stability that is significantly worse for a large number of entrained oscillators. The size of the source is defined precisely in terms of the participation ratio of the maximum energy localized state of the corresponding Anderson problem. For the system in Figs.~\ref{Fig: general},\ref{Fig: general qe0}, $p\simeq 9.36$. In this example, the frequency precision would not be improved by increasing the number of oscillators beyond about ten.

From Eq.~(\ref{Linear}) it is clear that the noise reduction factor $F/f$ depends on the parameters of the model only through the product $\gamma\sigma$, for the approximation to $\Gamma(\phi)$ used. Within this approximation, the scaling found from numerical solutions of Eq.~(\ref{Linear}) for a one dimensional lattice is shown in Fig.~\ref{Fig: general qe0}. The calculations were done for chains of length 100-1000, and the results were insensitive to the length providing it is much larger than the width of the localized state. A power law $F/f\propto (\gamma\sigma)^{0.6}$ is a good fit to the calculated results over the range considered.


\subsubsection{d-dimensional lattice}

\begin{figure}[tbh]
\begin{center}
\includegraphics[width=0.5\columnwidth]{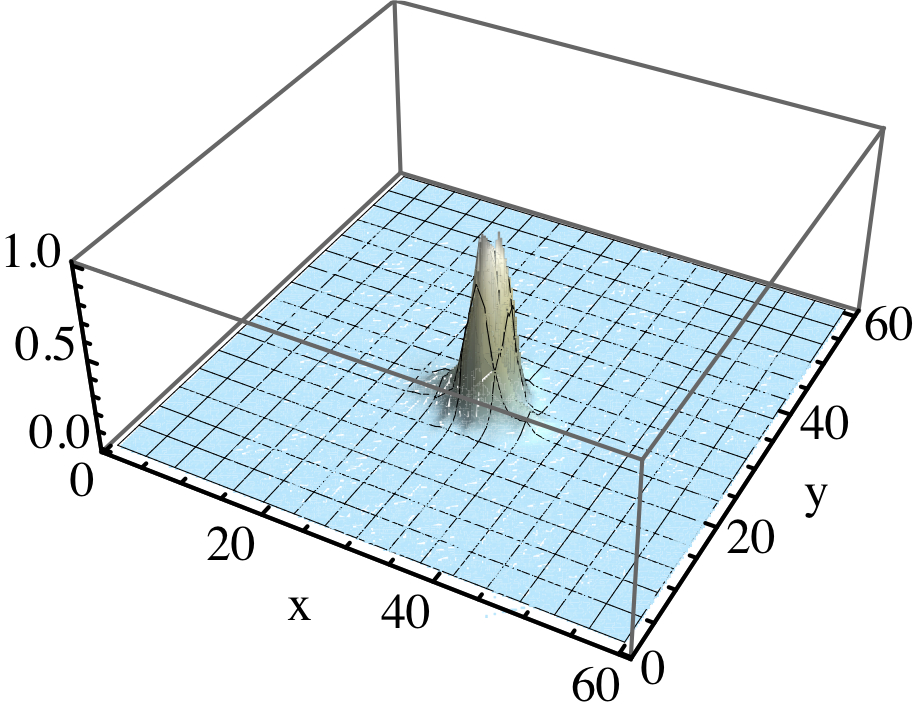}
\hspace{0.02\columnwidth}
\includegraphics[width=0.37\columnwidth]{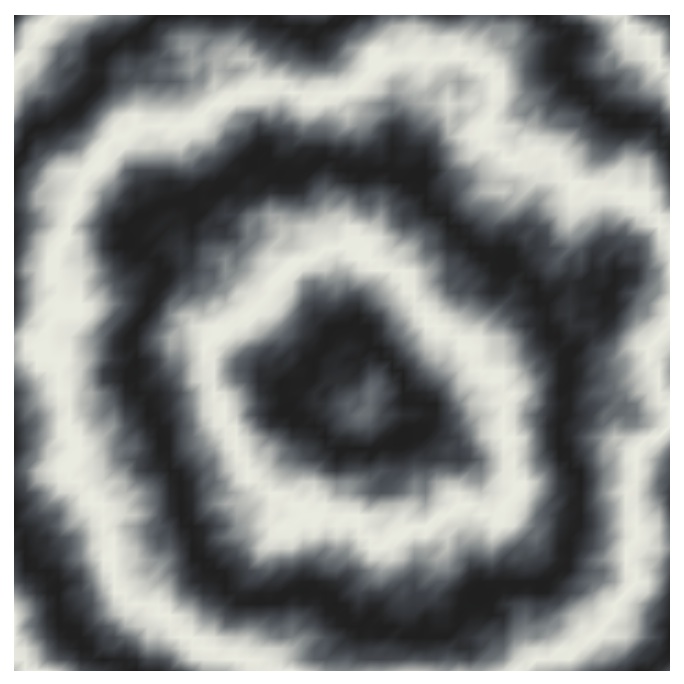}
\caption{\label{Fig: 2d}Source and waves for a 60 x 60 two dimensional lattice of oscillators with $\gamma=1,\sigma=2$. Left panel: zero-eigenvalue adjoint eigenvector $\mathbf e^{\ \dagger}_{0}$ showing the effective size of the source of the waves in the entrained state; right panel: grey scale plot of  $\cos\theta_{i,j}^{(s)}$ giving a snapshot of the waves emanating from the source.}
\end{center}
\end{figure}

The same argument applies to general dimension, although the structure of the Jacobean matrix is no longer tridiagonal. Choose any convenient labeling of the oscillators $\theta_{i},i=1\ldots N$. Nearest neighbor oscillators will not in general be adjacent in the list. However the Jacobean and its adjoint are still defined by
\begin{align}
J_{ij}&=q_{j}^{\text{max}}/q_{i}^{\text{max}},\quad\text{for $ij$ nearest neighbors}\\
J^{\ \dagger}_{ij}&=q_i^{\text{max}}/q_{j}^{\text{max}},\quad\text{for $ij$ nearest neighbors}\\
J^{\ \dagger}_{ii}&=J_{ii}=-\sum_{\text{nn }j}q_{j}^{\text{max}}/q_{i}^{\text{max}},
\end{align}
with other elements zero. The eigenvectors $\mathbf e_{0},\mathbf e_{0}^{\ \dagger}$ are as before, and the expression Eq.~(\ref{F/f}) for $F$ in terms of the inverse participation ratio is unchanged.
Thus I expect the noise reduction factor to scale as $r^{-d}$ with $r\sim p^{1/d}$ the radius of the maximum energy localized state in the $d$-dimensional Anderson localization problem.

Figure \ref{Fig: 2d} shows an example of a target wave entrained state for a 60 x 60 two dimensional lattice. The left panel is the adjoint eigenvector defining the source: the inverse participation ratio, yielding the improvement in the frequency precision, is 54.1 (cf.\ $N=3600$). The right panel is a plot of $\cos\theta_{ij}^{(s)}$, calculated from the Cole-Hopf transformation and the eigenvector $\mathbf{q}^{\text{max}}$, giving a snapshot of the waves in the entrained state.

\subsection{More general systems}
The results Eq.~(\ref{Noise reduction}) and Eq.~(\ref{adjoint}) remain valid for a more general coupling yielding the equations for the phase dynamics
\begin{equation}
\dot\theta_{i}=\omega_{i}+\sum_{j}K_{ij}\Gamma(\theta_{j}-\theta_{i})+\xi_{i}(t),\qquad i=1,\ldots,N.\label{Phase dynamics general}
\end{equation}
with $K_{ij}$ a symmetric matrix giving the strength of the coupling between oscillators $i$ and $j$. The small phase difference condition so that Eq.~(\ref{Small phase difference coupling}) may be used is now that the phase difference between \emph{any} two oscillators with nonzero $K_{ij}$ be small in the entrained state. The same analysis leads to the relationship Eq.~(\ref{F/f}) between the noise reduction factor and the participation ratio of the largest $E$ eigenvector $\mathbf q^{\text{max}}$ of the corresponding linear problem
\begin{equation}
Eq_{i}=\gamma\omega_{i}q_{i}+\sum_{j}K_{ij}(q_{j}-q_{i}).\label{Linear general coupling}
\end{equation}
Note that although the \emph{strength} of the coupling can be different for different paris of oscillators, the \emph{form} of the coupling Eq.~(\ref{General coupling}), and in particular the ratio of reactive to dissipative components, must be the same for this simple analysis to apply.

One generalization Eq.~(\ref{Phase dynamics general}) allows is to a lattice of oscillators with short range, but not just nearest neighbor, interactions. The scaling of the noise reduction with $\gamma\sigma$ will be the same as for nearest neighbor interactions, since the scaling properties of the Anderson problem are the same for these two cases. More generally, the method reduces the problem of calculating the improved frequency precision in the entrained state of a complex network of oscillators to solving the linear problem Eq.~(\ref{Linear general coupling}) for the network architecture and coupling parameters $K_{ij}$.

\section{Discussion}

The major result of this paper is that for oscillators on a lattice with short range coupling including a reactive component, the improvement of the frequency precision due to synchronization is limited to a factor given by the number of oscillators in the core source region of the waves that form the entrained state, rather than  a factor equal to the total number of oscillators, as is the case for purely dissipative coupling. I showed this result explicitly for the phase reduction description, in the limit of small enough disorder, or strong enough coupling, so that the phase differences between interacting oscillators are small in the entrained state. The size of the core region is given by the extent of the localized ground state of the corresponding linear Anderson problem, onto which the nonlinear phase equation is mapped by a Cole-Hopf transformation. The precise relationship is Eq.~(\ref{F/f}) relating the reduction in the phase noise to the inverse participation ratio of the localized state. This relationship remains true for general networks of oscillators providing the small phase difference approximation Eq.~(\ref{Small phase difference coupling}) applies for all interacting pairs of oscillators, and reduces the calculation of the frequency precision to the corresponding linear Anderson problem on the network.

Within the small phase difference approximation, the entrained state of waves propagating from the localized source is the unique state at long times. However, for the phase equations with the full coupling function Eq.~(\ref{General coupling}) other states may result depending on the initial conditions. This is particularly evident for two dimensional lattices, where spiral states are seen in numerical simulations starting from particular initial conditions\cite{SakaguchiShinomoto88}. Due to the topological constraint of integral $2\pi$ phase winding around the center, such a structure survives at long times, unless the core migrates to an open boundary. A second consequence of the topological structure is that there are necessarily large phase differences between nearest neighbor phases in the core, so that the small phase difference approximation breaks down. In the spiral state all the oscillators are again entrained to a single frequency, that probably depends just on the oscillators in the core region. It would be interesting to extend the analysis of the present paper to these spiral states.

\appendix\section{Derivation of the phase equation}\label{Appendix}

In this appendix I give a brief derivation of the stochastic phase equation Eq.~(\ref{Stochastic phase}). This equation follows from the general results for limit cycles of refs.\ \onlinecite{DemirMehrotra00,GoldobinTeramae10}, but the derivation is simpler for the phase reduction description.
I start from Eq.~(\ref{Phase})
\begin{equation}
\dot\theta_{i}=\omega_{i}+\sum_{\text{nn }j}\Gamma(\theta_{j}-\theta_{i})+\eta\xi_{i}(t),\qquad i=1,\ldots,N,\label{Phase eta}
\end{equation}
introducing a perturbation parameter $\eta$ to label the small noise. I will develop a perturbation expansion in $\eta$, and set $\eta\to 1$ at the end of the calculation. I will expand to first order in $\eta$ to extract the phase diffusion: continuing to second order would be needed to find the Lamb shift like correction to the mean frequency.

At zeroth order in $\eta$ the solution is the no-noise solution Eqs.~(\ref{Limit cycle}-\ref{Fixed point}), with $\Theta_{0}$ an arbitrary constant. In the presence of order $\eta$ noise I expect this overall phase to evolve on a slow time scale $T=\eta t$, $\Theta_{0}=\Theta_{0}(T)$, so that the phases will be given up to order $\eta$ by
\begin{equation}
\theta_{i}(t)=\theta_{i}^{(s)}+\Omega t+\Theta_{0}(T)+\eta\theta_{i}^{(1)}(t)+\ldots,
\end{equation}
with $\bm\theta^{(1)}(t)$ a correction to be found. Expanding the equation of motion up to order $\eta$ the equation for this correction is
\begin{equation}
\dot\theta_{i}^{(1)}-\sum_{j}J_{ij}\theta_{j}^{(1)}=-(\Theta'_{0}e_{0,i}-\xi_{i}(t)),\label{theta_1}
\end{equation}
with $J_{ij}$ the Jacobean Eq.~(\ref{Jacobean}), $\Theta'_{0}=d\Theta_{0}/dT$, and $\mathbf e_{0}=(1,1,1\ldots 1)$. Components of $\bm\theta^{(1)}$ along eigenvectors of $\mathbf J$ with negative eigenvalues will have some finite value given by inverting this equation. However, for the component along the zero-eigenvalue eigenvector, there is no restoring force, and any nonzero value of the right hand side will lead to large values of $\theta_{i}^{(1)}$ at large times, violating the assumption that $\bm\theta^{(1)}$ gives a small correction. This component is extracted by multiplying on the left by the adjoint eigenvector $\mathbf e^{\ \dagger}_{0}$ since $\mathbf e^{\ \dagger}_{0}\cdot\mathbf J=0$. This leads to the solvability condition for $\bm\theta^{(1)}$ to remain finite
\begin{equation}
\Theta'_{0}=\frac{\mathbf e^{\ \dagger}_{0}\cdot\mathbf\xi}{\mathbf e^{\ \dagger}_{0}\cdot\mathbf e_{0}}.
\end{equation}
Returning to the original variables and setting $\eta\to 1$ gives the stochastic equation for the overall phase
\begin{equation}
\dot\Theta(t)=\Omega+\Xi(t),
\end{equation}
with the correlation function of the effective noise
\begin{equation}
\langle\Xi(t)\Xi(t')\rangle=\frac{\sum_{ij}e^{\ \dagger}_{0,i}e^{\ \dagger}_{0,j}\langle\xi_{i}(t)\xi_{j}(t')\rangle}{(\mathbf e^{\ \dagger}_{0}\cdot\mathbf e_{0})^{2}},
\end{equation}
giving Eq.~(\ref{Phase noise}) with Eq.~(\ref{Noise reduction}) for equal, uncorrelated white noise of strength $f$ acting on each individual oscillator. Note that the result does not depend on the choice of normalization for $\mathbf e^{\ \dagger}_{0}$. A specific normalization choice for $\mathbf e_{0}$ was made in setting up Eq.~(\ref{theta_1}).

\begin{acknowledgments}
This work was supported by NSF grant number DMR-1003337. I thank Eyal Kenig and Tony Lee for comments on the manuscript.
\end{acknowledgments}

\bibliography{Bibliography}

\end{document}